\documentstyle[prl,aps,multicol,epsf]{revtex}
\newcommand{\beq}{\begin{equation}}
\newcommand{\eeq}{\end{equation}}

\begin{document}

\title{Raman scattering from a superconductivity-induced bound state in
$MgB_2$}

\author{R. Zeyher}

\address{Max-Planck-Institut\  f\"ur\
Festk\"orperforschung,\\ Heisenbergstr.1, 70569 Stuttgart, Germany }

\date{\today}


\maketitle

\begin{abstract} 
It is shown that the sharp peak in the $E_{2g}$ Raman spectrum of
superconducting $MgB_2$ is due to a bound state caused by the 
electron-phonon coupling. Our theory explains why this peak appears only
in the spectra with $E_{2g}$ symmetry and only in the $\sigma$ but not
$\pi$ bands. The properties of the bound state and the 
Raman spectrum are investigated, also in the presence of impurity 
scattering.
 
\end{abstract}

\begin{multicols}{2}


Electronic Raman scattering in superconductors probes the density and, to
some extent, the momentum dependence of quasi-particle excitations
across the gap. As a result the spectra are usually dominated by broad pair 
breaking peaks reflecting the magnitude and anisotropy of the 
gap as well as scattering processes of the quasi-particles.
Experimental spectra in the two-gap superconductor $MgB_2$ show the typical
features of a dirty s-wave superconductor\cite{Quilty1,Quilty2}. 
A new and surprising feature in the $E_{2g}$ spectrum is a sharp peak
near the larger of the two gaps which in our opinion cannot be explained
as a pair breaking peak because of the large impurity scattering rate of
quasi-particles in the present $MgB_2$ samples. 
Below we will argue that this peak is the analogue
of the magnetic resonance peak observed in high-T$_c$ cuprates\cite{Levin}
where the electron-phonon interaction takes over the role of the Heisenberg 
interaction and strongly scatters the excitations across the gap.

Neglecting the momenta of the incident and scattered photons the
differential Raman cross section can be written as 
\begin{equation}
{{d^2R}\over{d\omega d \Omega}} = -{{e^4}\over {\pi c^4}} (1+n(\omega))
\chi''(\omega).
\label{cross}
\end{equation}
$e$ is the charge of an electron, $n(\omega)$ 
the Bose distribution function, $\omega$ the difference in frequency between 
incident and scattered light, and $\chi''$ the imaginary part of a retarded
Green's function $\chi$ at zero momentum. The corresponding Matsubara
function is 
\begin{equation}
\chi(i\omega_n) = -\int_0^\beta d \tau e^{i\omega_n \tau} \langle
T_{\tau} \rho(\tau) \rho(0) \rangle,
\label{chi}
\end{equation}
where $\beta$ is the inverse temperature $1/T$, $\omega_n = 2n \pi T$
are bosonic Matusbara frequencies, $T_{\tau}$ the time ordering operator, and 
$\rho$ an effective density operator with momentum zero. In the following we
will be interested in spectra related to the $E_{2g}$ symmetry of
$D_{6h}$, the point group of $MgB_2$. $\rho$ has then the form
\beq
\rho = \sum_{{\bf k}\sigma n} \gamma_n({\bf k})
c^\dagger_{\sigma n}({\bf k}) c_{\sigma n}({\bf k}) +R(b + b^\dagger),
\eeq 
\beq
\gamma_n({\bf k}) = 
\Bigl({{{\partial ^2} \epsilon_n({\bf k})}
\over{\partial k_x^2}} - {{{\partial ^2} \epsilon_n({\bf k})}
\over{\partial k_y^2}}\Bigr). 
\label{gamma}
\eeq
$c^\dagger_{\sigma n}
(\bf k)$ and $c_{\sigma n}(\bf k)$ are creation and annihilation 
operators for electrons with momentum ${\bf k}$, spin direction $\sigma$, 
band label $n$, and energy $\epsilon_n({\bf k})$. $MgB_2$ has 
only one Raman-active ${\bf q}=0$ phonon, and this phonon has 
$E_{2g}$ symmetry. Its frequency is denoted
by $\Omega$, its creation and annihilation operators by $b^\dagger$ and $b$,
respectively, and the corresponding element of the phononic Raman tensor
by $R$. Without loss of generality we assume that this phonon 
transforms as the first basis vector of the two-dimensional $E_{2g}$
representation in accordance with Eq.(\ref{gamma}).

For the evaluation of $\chi$ we use the Hamiltonian $H = H_0 + H'$ with
\begin{eqnarray}
H_0 &=& \sum_{{\bf k}\sigma n} \epsilon_n({\bf k}) 
c^\dagger_{\sigma n}({\bf k})
c_{\sigma n}({\bf k})  
+\sum_{{\bf q}j} \Omega_j({\bf q}) (b^\dagger_j({\bf q}) b_j({\bf q}) + 
{1\over{2}})
\nonumber\\ 
&-&\sum_n \Bigl( \Delta_n c^\dagger_{\uparrow n}
({\bf k}) c^\dagger_{\downarrow n}(-{\bf k}) +h .c.\Bigr),
\label{H_0}
\end{eqnarray}
\begin{eqnarray}
H' = \sum_{{\bf k}{\bf q}n}g_{nj}({\bf k}{\bf q}) c^\dagger_{\sigma n}
({\bf k}+{\bf q}) c_{\sigma n}({\bf k}) (b_j({\bf q}) + b^\dagger_j(-{\bf q}))
\nonumber\\
+\sum_{{{{\bf k}{\bf q} \sigma}}\atop{nn'}} V_{nn'}({\bf k}{\bf q}) 
c^\dagger_{\sigma n}({\bf k}) c_{\sigma n'}({\bf q}).
\label{H'}
\end{eqnarray}
$\Delta_n$ is the gap parameter for s-wave superconductivity
in the band $n$, $\Omega_j({\bf q})$ the frequency of a general
phonon with momentum $\bf q$ and branch label $j$, $g$ the coupling
constant for intraband electron-phonon scattering and $V$  
a random potential for intraband impurity scattering. Interband
phonon scattering can be neglected in $H'$ because only zero momentum
transfers occur in the approximation used below.

In a first step we perform an infinite summation over bubble diagrams
by introducing the irreducible Green's function $\tilde{\chi}$.
$\tilde{\chi}$ contains all diagrams to $\chi$ which cannot be decomposed
into two parts by cutting one phonon line. The average over impurities
yields impuritiy lines with similar properties as phonon lines. However,
bubble diagrams connected by impurity
lines are not possible in this case so that the above definition of 
irreducibility is appropriate.
Analytically, one obtains
\beq
\chi = \tilde{\chi}_{11} +(
\tilde{\chi}_{12}
+R) D (R+\tilde{\chi}_{21}),
\label{chi1}
\eeq
\beq
D = D^{(0)} + D^{(0)} \tilde{\chi}_{22} D.
\label{D}
\eeq
The omitted frequency and momentum arguments in the Green's functions in 
Eqs.(\ref{chi1}) and (\ref{D}) are $i\omega_n$ and $0$,
respectively. $\tilde{\chi}_{11}$ denotes the irreducible
Green's function associated with the two vertices 
$\gamma_n({\bf k})$. 
Similarly, the vertices in the functions $\tilde{\chi}_{12}$
and $\tilde{\chi}_{22}$ 
are $\gamma_n({\bf k})$ and $g_n({\bf k}0)$, and two times 
$g_n({\bf  k}0)$, respectively.
The free phonon progator $D^{(0)}$ is given by 
$-{2\Omega/({\omega_n^2+\Omega^2})}$.

A sensible approximation for the evaluation of $\tilde{\chi}$ is the
ladder approximation plus the corresponding self-energy corrections
where the interaction lines are due to phonons
or impurities. Only that part of the interaction can contribute 
in the ladder diagrams which transforms in the same way as the vertices
$\gamma$ and $g$. This means
in our case that only the $E_{2g}$ component of the phonon-mediated 
interaction, which usually is considered to be neglegible in a s-wave
superconductor, would enter.  
We therefore will evaluate $\tilde{\chi}$ only in the presence of
random impurities using the Born approximation and the dirty limit
for each band. Assuming that the vertices and the
interaction can be evaluated right on the Fermi surface, expanding
$\gamma$ and $g$ in terms of Fermi surface harmonics\cite{Allen}
of $E_{2g}$ symmetry, ${\Phi^{(n)}_L}({\bf k})$, normalized as in Eq.(1) of
Ref.\cite{Allen}, 
\beq
\gamma_n({\bf k}) = \sum_L \alpha^{(n)}_{1L} \Phi^{(n)}_L({\bf k}),
\eeq
\beq
g_n({\bf k}0) = \sum_L \alpha^{(n)}_{2L} \Phi^{(n)}_L({\bf k}),
\eeq
and assuming that the interaction is diagonal in $L$ we obtain
\beq
\tilde{\chi}_{ij}(i\omega_n) = 
\sum_{L n} \alpha^{(n)}_{iL} \alpha^{(n)}_{jL} N_F^{(n)}
\tilde{\chi}(i \omega_n,\Delta_{n},\tau_n^{-1}).
\eeq
$N_F^{(n)}$ is the density of states at the
Fermi surface for one spin direction due to the band $n$ and $1/\tau_n$ an
effective scattering rate. 
After an analytic continuation $i\omega_n \rightarrow \omega + i\eta$
the imaginary part of the Green's function
$\tilde{\chi}$, which is independent of $L$,
becomes\cite{Devereaux}
\begin{eqnarray}
Im \;\tilde{\chi}(\omega,\Delta, \tau^{-1}) =-{{8\Delta \omega}\over
{\omega+2\Delta}} \Theta(\omega-2\Delta){{\tau^{-1}}\over{\omega^2+\tau^{-2}}}
\nonumber\\
\cdot \Bigl( {{(\omega-2\Delta)^2}\over{4\Delta \omega}} E(\alpha)
+ {{\tau^{-2} +\omega^2
    +4\Delta\omega}\over{\tau^{-2}+\omega^2+2\Delta\omega}}
F(\alpha) \nonumber\\
+{{8\Delta^2\omega^2}\over{(\omega^2+\tau^{-2})^2-4\Delta^2\omega^2}}
\Pi(N,\alpha) \Bigr),
\label{B}
\end{eqnarray}
with 
\beq
\alpha = (\omega-2\Delta)/(\omega +2\Delta),
\label{alpha}
\eeq
\beq
N = {{1+\alpha^2 \omega_1}\over{1+\omega_1}},
\eeq
\beq
\omega_1 = \tau^{-2}(1-4\Delta^2/(\omega^2+\tau^{-2}))/(\omega-2\Delta)^2.
\eeq
$\Theta$ is the theta function and $F$, $E$, and $\Pi$ are complete
elliptical integrals of the first, second, and third kinds, respectively.
The existence of only two gaps in the experimental
spectra\cite{Quilty1,Quilty2,Gonnelli} as well as theoretical
arguments\cite{Mazin}
suggest that the dirty limit applies even within the two-band complex
of $\sigma$ and $\pi$ bands (denoted by the index $\rho$ in the
following) and that the 
interband impurity scattering
between $\sigma$ and $\pi$ bands is neglegible. As a result $\Delta_n$
and $\tau_n^{-1}$ can be considered to be the same within the manifold of
$\sigma$ or $\pi$ bands.
Introducing then the effective couplings
\beq
\lambda^{(\rho)}_{ij} = \sum_{L\atop{n \in \rho}} \alpha^{(n)}_{iL} 
\alpha^{(n)}_{jL}N_F^{(n)},
\eeq
we obtain
\beq
\tilde{\chi}_{ij}(\omega) = \sum_{\rho=\sigma,\pi}
\lambda_{ij}^{(\rho)} \tilde{\chi}(\omega,\Delta_{\rho}, 
\tau^{-1}_{\rho}).
\eeq
Choosing for the first function $\Phi_1^{(n)}(\bf k)$
the properly normalized function $\sim \gamma_n(\bf k)$, the
summations over $L$ collapse to one term $L=1$ if either $i$ or $j$
is equal to 1.
In the clean limit $1/\tau \rightarrow 0$ Eq.(\ref{B}) reduces 
to the analytical formula Eq.(16) of Ref.(\cite{Zeyher1}).

Using the tight-binding fit to the band structure of Ref.\cite{Kong},
$eV$ as energy and the lattice constant $a$ as length units,
we find $N_F^{(1)} = 0.049$, $N_F^{(2)} = 0.109$, $N_F^{(3)} = 0.101$, 
$N_F^{(4)} = 0.104$, where $n=1,2$ denote the light and heavy $\sigma$
bands and $n=3,4$ the $\pi$ bands. In the case of $\pi$ bands we have
scaled the published tight-binding parameters slightly in order to reproduce
the correct densities. Furthermore, we obtain $\alpha_1^{(1)} = 0.101$, 
$\alpha_1^{(2)} = 0.261  $, $\alpha_1^{(3)} = 2.531  $, 
$\alpha_1^{(4)} = 2.376  $, and thus
$\lambda_{11}^{(\sigma)} = 0.008$, and $\lambda_{11}^{(\pi)} = 1.238$.
The rather small values for $\alpha_1^{(1)}$, $\alpha_1^{(2)}$, and 
$\lambda_{11}^{(\sigma)}$ reflect the fact that 
the $\sigma$ bands are rather isotropic in the ab-plane near their narrow
cylindrical Fermi
surfaces and thus cannot contribute much in the $E_{2g}$ channel.
As a consequence we may safely put $\lambda_{11}^{(\sigma)} = 
\lambda_{12}^{(\sigma)} = \lambda_{21}^{(\sigma)} = 0$. The
$\sigma$ contribution to $\tilde{\chi}$ arises then solely from
the coupling of the phonon to the light and its spectral features
are determined by the dimensionless electron-phonon coupling constant 
$\lambda = {2\over\Omega} \lambda_{22}^{(\sigma)}$ and the scattering rate
$1/\tau_\sigma$. 

We will first consider the $\sigma$ contribution to 
$\tilde{\chi}$ and drop the index $\sigma$ everywhere in order to simplify
the notation.
Fig.1 shows the real (dashed lines) and imaginary (solid lines) part
of $\tilde{\chi}$ for the gap $2\Delta = 110 cm^{-1}$ and two scattering
rates $1/\tau = 0$ and $1/\tau = 200 cm^{-1}$. In the clean case the imaginary
part $\tilde{\chi''}$ exhibits a square root singularity if $\omega$
approaches $2\Delta$ from above. Impurity scattering transforms this
singularity into a step at $2\Delta$ with height $2\pi\Delta \tau $, and 
produces for sufficiently strong scattering
a very broad minimum near $1/\tau$.
The real part $\tilde{\chi}'$ shows in the clean case a square root
singularity below $2\Delta$ and finite and positive values above $2\Delta$.
Taking impurity scattering into account $\tilde{\chi}'$ becomes much more
symmetric around $2\Delta$ compared to the clean case, 
\begin{figure}[h]
\centerline{
      \epsfysize=6cm
      \epsfxsize=7cm
      \epsffile{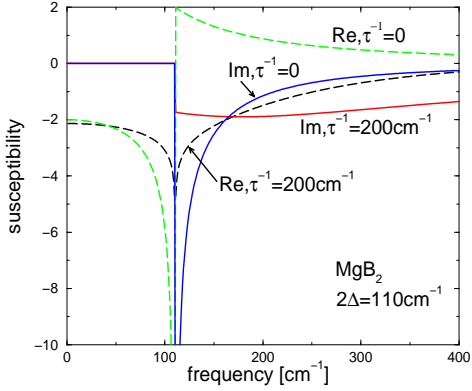}}
\label{fig1}
\caption 
{Real (dashed lines) and imaginary (solid lines) part of the susceptibility
$\tilde{\chi}$ for two impurity scattering rates $\tau^{-1}$ and parameters
appropriate for $MgB_2$.}
\end{figure}
\noindent
exhibiting there a logarithmic singularity and negative values over a large 
region above $2\Delta$.

\begin{figure}[h]
\centerline{
      \epsfysize=6cm
      \epsfxsize=7cm
      \epsffile{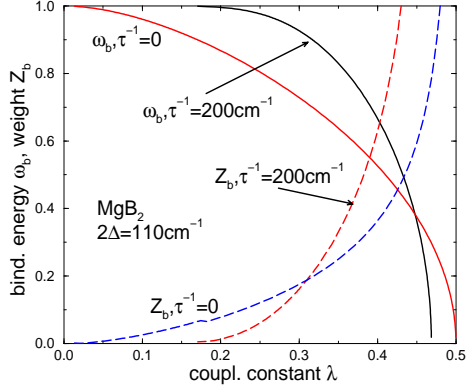}}
\label{fig2}
\caption 
{Energy $\omega_b/2\Delta$ (solid curves) and spectral weight $Z_b$ (dashed 
curves) of the bound state for two impurity scattering rates $\tau^{-1}$
as a function of the electron-phonon coupling constant $\lambda$.}
\end{figure}
\noindent

The curves in Fig. 1 suggest that the phonon Green's function $D$
will develop a bound state inside the gap\cite{Zeyher2}, 
and that this will occur for all
values for $\lambda$ and $1/\tau$. The frequency of the bound state,
$\omega_b$, is determined by the vanishing of the denominator of $D$,
i.e., by the equation,
\beq
\omega^2_b = \Omega^2 + \Omega^2 \lambda \tilde{\chi}'(\omega_b,
\Delta, 1/\tau).
\label{bound}
\eeq 
Expanding the denominator of $D$ around $\omega_b$ one finds for the
spectral weight $Z_b$ 
\beq
Z_b = {{2 \Omega}\over{2\omega_b - \lambda \Omega^2 \partial { \tilde{\chi}'}
(\omega_b,\Delta,1/\tau)/\partial \omega_b}}.
\label{Z}
\eeq

Fig.2 shows $\omega_b$ and $Z_b$ as a function of $\lambda$ for the two
scattering rates $1/\tau = 0$ and $1/\tau = 200 cm^{-1}$. $\omega_b$
approaches zero at $\lambda_{CDW}=0.50$ and $0.47$, respectively. This means
that for $\lambda > \lambda_{CDW}$ the superconducting state is unstable
against the formation of a charge density wave with $E_{2g}$
symmetry. For $\lambda < \lambda_{CDW}$ $\omega_b$ and $Z_b$ increase and
decrease rapidly with decreasing $\lambda$ approaching their limiting 
values $1$ and $0$ in an exponential (for $1/\tau \neq 0$) or powerlawlike
(for $1/\tau = 0$) manner.

One peculiar feature of $D$ is that any background imaginary part in its
self-energy, for instance due to $\pi$ electrons,
will be diminuished by the factor $1/Z_b$ near $2\Delta$.
This means that the bound state will not be broadened
\begin{figure}[h]
\centerline{
      \epsfysize=6cm
      \epsfxsize=7cm
      \epsffile{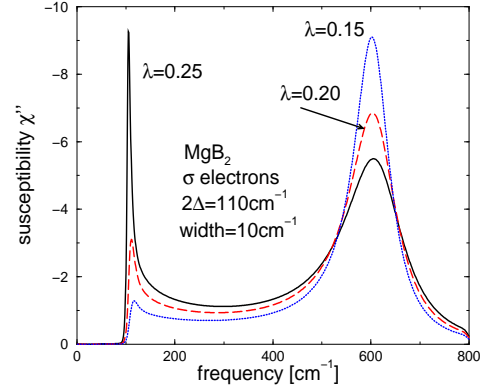}}
\label{fig3}
\caption 
{Susceptibility $\chi''$ of $\sigma$ electrons at $T=0$
for the scattering rate $\tau^{-1}=200 cm^{-1}$.}
\end{figure}
\noindent
and disappear in $\chi''$
for small $\lambda$'s, even in the presence of strong impurity 
scattering, but instead will sharpen and loose weight
when $\omega_b$ approaches 1. To make the situation more realistic one
should allow for broadening
effects due to inhomogenieties etc. which may be taken into account by
folding $\tilde{\chi}'$ with a Gaussian with a width 
$\delta$\cite{Quilty2,Devereaux}. Fig.3
shows $\chi''$ for three different $\lambda$ and a width of 
$\delta = 10 cm^{-1}$. The employed large scattering rate 
$\tau^{-1}=200 cm^{-1}$ has wiped out completely the usual pair breaking
peak due to the square root divergence of $\chi''$ at $2\Delta$. On the
other hand, the electron-phonon coupling accumulates spectral weight
near $2\Delta$ and produces a pronounced bound state inside the
gap at larger couplings.

Figs.2 and 3 allow to estimate a realistic value for $\lambda$. Identifying
$\omega_b$ with the observed sharp peak in the $E_{2g}$ spectrum we have
$\omega_b = 104 \pm 1 \;cm^{-1}$ whereas a recent tunneling 
experiment\cite{Gonnelli} gave for
the gap $2\Delta = 114 \pm 6 \;cm^{-1}$. This suggest that $\omega_b$ is
different from $2\Delta$, i.e., the sharp peak should not be identified 
with the gap, and the ratio $\omega_b/2\Delta$ is $0.91 \pm 0.06$.
Fig.2 indicates then that $\lambda$ must be smaller than $0.3$. Comparing
the intensities of the bound state and the phonon line in Fig.3 with the
experimental curve\cite{Quilty1} one finds that $\lambda$ must be near the
interval between 0.15 and 0.20. These values are substantially smaller than 
the value 0.38 obtained from band structure calculations\cite{Liu}.

Additional evidence for the importance of the electron-phonon coupling
in the $E_{2g}$ spectrum comes from the experimental result that the two 
$A_{1g}$ and the $E_{1g}$ spectra do not show any pronounced peak near 
$2\Delta$. We explain this by the fact that, according to group theory,
$MgB_2$ has no ${\bf q}=0$ phonons with such symmetries so that no

\begin{figure}[h]
\centerline{
      \epsfysize=5cm
      \epsfxsize=8cm
      \epsffile{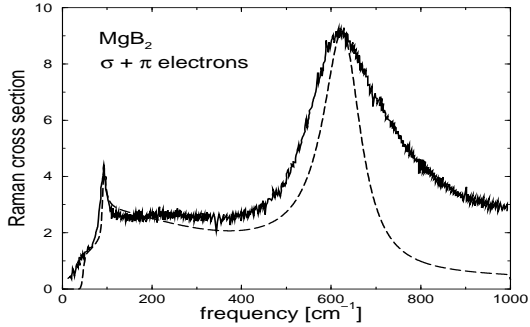}}
\label{fig4}
\caption 
{Experimental (solid line) and theoretical (dashed line) 
$E_{2g}$ Raman cross section.}
\end{figure}
\noindent
bound states can be formed. Further support for the
present theory comes from the absence of a
peak near the gap of about 47 $\;cm^{-1}$ in the  $\pi$ bands.
LDA calculations show that the coupling of $\pi$ electrons to the $E_{2g}$
phonon is by a factor 3 or 4 smaller than for
$\sigma$ electrons\cite{Liu}. We find that the bound state
structure near the $\pi$ gap is invisible for such a coupling
which explains its absence in the
experimental spectra. 
Using $\lambda^{(\pi)}_{12} = \lambda^{(\pi)}_{21} = 0$,
$\lambda^{(\pi)}_{22}/\lambda^{(\sigma)}_{22}=1/3$,
$\lambda^{(\pi)}_{11}/R^2 = 1/400$, $\lambda=0.16$,
$\Omega = 620 cm^{-1}$, $\delta=5cm^{-1}$, a $\pi$ gap of $43 cm^{-1}$
and a somewhat reduced $\sigma$ gap of $96.5\;cm^{-1}$, accounting for
the finite temperature of the experimental data, 
the resulting Raman cross section is shown as the dashed line in Fig.4, 
together
with the experimental $E_{2g}$ spectrum\cite{Quilty1}.
The dashed line reproduces well the main features of the experimental curve,
especially at small frequencies.
The quantitative discrepancy in the phonon region suggests that only part
of the phonon broadening is caused by the electron-phonon coupling and that
anharmonicity and deviations from the assumed constant density of states
may play a role. 
The theoretical superconductivity-induced hardening of the
phonon frequency in Fig.4 is 5 $cm^{-1}$ and thus somewhat smaller than 
the experimental value\cite{Quilty1} of $7-10 \;cm^{-1}$.  

In conclusion, we have shown that the $E_{2g}$ spectrum in superconducting
$MgB_2$ can be understood as a superposition of a phonon line coupled
strongly to $\sigma$ electrons creating hereby a bound state in the
gap, and a background due to rather uncorrelated $\pi$ electrons. This means 
that $MgB_2$ is to our knowledge the first s-wave superconductor 
where a bound state in the superconducting gap due to residual interactions
has been observed and identified. The obtained electron-phonon coupling
constant $\lambda \sim 0.2$ for $\sigma$ electrons is only half of the
band structure value, a discrepancy, which presently is not well understood.
 
The author thanks O. Dolgov, J. Kortus and I. Mazin for useful discussions.

\end{multicols}
\end{document}